\documentclass[final,3p,times,twocolumn]{elsarticle}

\biboptions{sort&compress} 
\usepackage{graphicx} 
\usepackage{dcolumn}
\usepackage{bm}
\usepackage{float}
\usepackage{comment} 
\usepackage{amssymb}
\usepackage{amsmath}

\usepackage{booktabs} 
\usepackage{xcolor} 

\usepackage[version=4]{mhchem}

\DeclareUnicodeCharacter{2212}{-} 
\journal{Journal Name Here}

\begin{document}
\raggedbottom        
\sloppy              
\emergencystretch=2em 

\begin{frontmatter} 

\title{Investigating structure and physical properties of quaternary layered transition metal oxide Na$_2$Cu$_2$TeO$_6$ }

\author[1,2]{Shubham Patil}
\author[1,2]{S. D. Kaushik\corref{cor1}}
\ead{sdkaushik@csr.res.in}
\cortext[cor1]{Corresponding author}

\affiliation[1]{organization={UGC-DAE Consortium for Scientific Research, Mumbai Centre},
            addressline={Bhabha Atomic Research Centre},
            city={Mumbai},
            postcode={400 085},
            state={Maharashtra},
            country={India.}}
 \affiliation[2]{organization={Savitribai Pule Pune University},
            addressline={Ganeshkhind Road},
            city={Pune},
            postcode={411 007},
            state={Maharashtra},
            country={India.}}           

\date{\today}

\begin{abstract}
We investigated the crystal structure, magnetic behavior, and optical properties of the layered honeycomb compound Na$_2$Cu$_2$TeO$_6$. X-ray and neutron diffraction confirmed a monoclinic \(C2/m\) structure, with Cu$^{2+}$ ions arranged in dimerized chains. Magnetic susceptibility measurements yielded a Curie-Weiss temperature of \(\Theta_{\text{CW}} = -212.5\) K and an effective moment \(\mu_{\text{eff}} = 0.16\,\mu_B\), significantly lower than the expected spin-only value, indicating the presence of strong antiferromagnetic interactions and enhanced quantum fluctuations. A broad maximum near 160 K in the susceptibility data is consistent with short-range one-dimensional antiferromagnetic correlations. Magnetization measurements showed negligible coercivity, and specific heat data revealed no anomalies down to 3 K. Temperature-dependent neutron diffraction showed no evidence of long-range magnetic order. Optical absorption studies using UV-Visible spectroscopy displayed a sharp absorption edge in the UV region. Tauc analysis estimated a direct optical band gap of approximately 2.10 eV, with no clear indication of an indirect transition. These observations provide insight into the interplay between structural distortions, low-dimensional magnetism, and optical behavior in Na$_2$Cu$_2$TeO$_6$.

\end{abstract}
\end{frontmatter}

\section{Introduction}

The study of novel quantum states in low-dimensional spin systems, particularly those based on honeycomb lattices of magnetic atoms, has garnered significant attention due to their potential for exhibiting unconventional electronic and magnetic properties. These systems host a range of fascinating phenomena, such as spin liquids, superconductivity, and topological phase transitions, arising from intricate magnetic interactions. A key focus of recent research has been to search for compounds exhibiting spin liquid state as predicted by Kitaev in materials with 4d or 5d electrons, due to their potential applications in advanced technological devices~\cite{kitaev2006anyons,matsuda2010disordered,heinrich2003potential}. This exotic quantum state emerges due to strong spin-orbit coupling and bond-directional (anisotropic) interactions, as outlined in the Kitaev model, which describes anisotropic interactions between spin-1/2 particles on a honeycomb lattice~\cite{PhysRevLett.102.017205}.  

While Kitaev materials have been predominantly explored in 4d/5d transition metals, 3d-based honeycomb magnets offer an alternative route where frustration arises from competing exchange interactions rather than spin-orbit coupling. Na$_2$Cu$_2$TeO$_6$ is an example of such a system, where Cu$^{2+}$ ions form a distorted honeycomb lattice. This compound belongs to a broader class of layered honeycomb oxides that allow various substitutions, such as Na-Li, Sb-Bi, and Cu-Co-Ni, for tuning their magnetic interactions~\cite{skakle1997synthesis,zvereva2015zigzag,zvereva2012monoclinic,berthelot2012studies,berthelot2012new,schmidt2013solid,sankar2014crystal,seibel2013structure}.  

These honeycomb-layered oxides, characterized by compositions A$_2$M$_2$DO$_6$ and A$_3$M$_2$DO$_6$, where A represents alkali or coinage metals (Li, Na, K), M is a transition metal (Ni, Co, Zn, Cu), and D is a chalcogen or pnictogen metal (Te, Sb, Bi)~\cite{Kanyolo2023-yn}, provide a versatile platform to investigate the role of competing interactions (J$_2$/J$_1$ and J$_3$/J$_1$) in stabilizing novel ground states. The presence of edge-sharing MO$_6$ octahedra enables higher-order superexchange interactions beyond nearest neighbors. Furthermore, interlayer Na/Li/Ag ions modulate the magnetic coupling strength, contributing to the structural softness of these materials. Such properties also make these compounds attractive for applications as cathode materials in rechargeable batteries~\cite{roudebush2013structure,derakhshan2007electronic,xu2005synthesis,miura2006spin,zvereva2012monoclinic,berthelot2012new,evstigneeva2011new,schmitt2014microscopic,bera2017zigzag,zvereva2015magnetic,wong2016zig,lefranccois2016magnetic,zvereva2016orbitally}.  

\begin{figure*}[htbp]
	\centering 
	\includegraphics[width=0.93\textwidth, angle= 0]{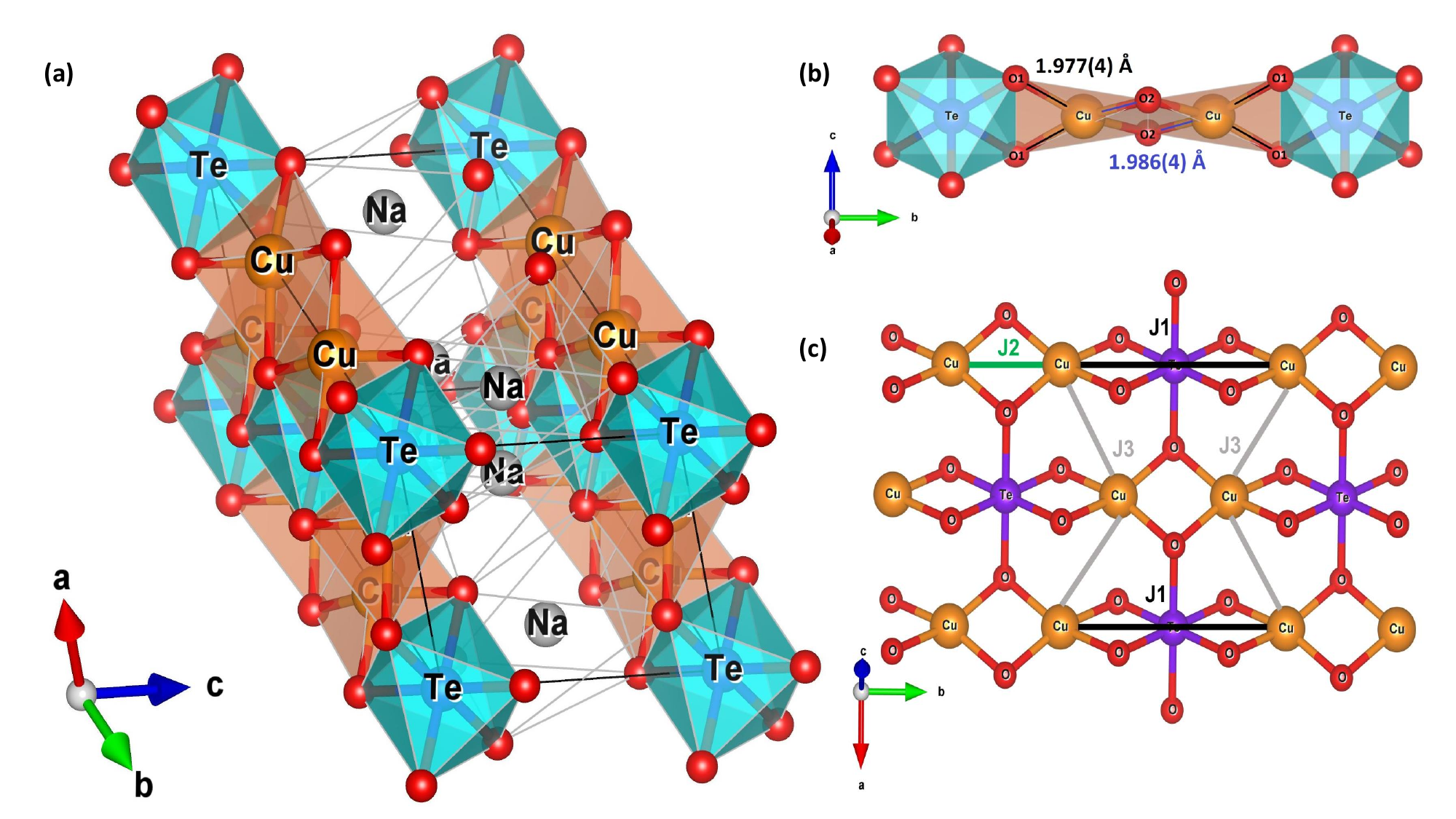}	
	\caption{Crystal structure of Na$_2$Cu$_2$TeO$_6$ determined from neutron diffraction analysis. (a) 3D view of the crystal structure where Na ions reside between Cu$_2$TeO$_6$ layers. (b) Planar arrangement of CuO$_4$ plaquettes showing distinct Cu-O$_1$ and Cu-O$_2$ bond lengths in the $bc$ plane. (c) Crystal structure along the $c^{\ast}$ axis, where Cu$^{2+}$ ions form chains along the $b$ axis. The Cu$^{2+}$ spins interact via alternating J$_1$ (black) and J$_2$ (green) couplings along the chain, and J$_3$ (light gray) interactions between chains. Visualization using VESTA software~\cite{Momma2011-VESTA}.}
	\label{fig_CrystalStructure}%
\end{figure*}

Recent studies on related honeycomb compounds, such as Na$_3$Co$_2$SbO$_6$ and Na$_2$Co$_2$TeO$_6$, have demonstrated the potential of earth-abundant elements (Co, Ni) to realize Kitaev-like spin-liquid behavior, as opposed to rare elements like Ru, Rh, or Ir. The monoclinic polymorph of Na$_2$Co$_2$TeO$_6$ exhibits a single antiferromagnetic transition, suggesting enhanced frustration due to out-of-plane spin canting compared to its hexagonal counterpart~\cite{PhysRevB.108.064405,wong2016zig,viciu2007structure,yan2019magnetic,bera2017zigzag,xiao2019crystal,lin2021field,mukherjee2022ferroelectric,yao2020ferrimagnetism}. Additionally, disorder and frustration effects have been linked to the absence of long-range magnetic ordering in compounds such as Na$_3$LiFeSbO$_6$, Na$_4$FeSbO$_6$, and Li$_4$MnSbO$_6$~\cite{schmidt2014synthesis,bhardwaj2014evidence}.  

Density functional theory (DFT) calculations suggest long-range zigzag antiferromagnetic ordering in (Li/Na)$_3$Ni$_2$SbO$_6$~\cite{zvereva2015zigzag}, a feature experimentally observed in Na$_3$Co$_2$SbO$_6$~\cite{wong2016zig}. Furthermore, low-temperature AFM ordering has been reported in O3-type honeycomb compounds such as Na$_3$M$_3$SbO$_6$ (M = Cu, Ni)~\cite{schmidt2013solid,roudebush2013structure,viciu2007structure}, Na$_3$Ni$_2$BiO$_6$~\cite{seibel2013structure}, Li$_3$Ni$_2$SbO$_6$~\cite{zvereva2012monoclinic}, and P2-type Na$_2$M$_2$TeO$_6$ (M = Co, Ni)~\cite{berthelot2012studies,schmidt2013solid,viciu2007structure}. However, the true quantum ground states of these materials remain unclear, requiring further experimental investigation. The impact of intermediate Na/Li layers on interlayer magnetic correlations is also an open question.

Quantum spin systems with $S = 1/2$, such as cuprates, provide a valuable platform for studying low-dimensional magnetism. Their behavior is governed by the Cu 3d$^9$ electron configuration, combined with geometric frustration and competing exchange interactions. These effects enhance quantum fluctuations and destabilize classical magnetic order. An example is the $S = 1/2$ Heisenberg compound Na$_2$Cu$_2$TeO$_6$, where the interplay between lattice distortions and electronic interactions plays a crucial role~\cite{schmitt2014microscopic}.  

Although the magnetic and crystal structures of Na$_2$Cu$_2$TeO$_6$ (NCTO) have been previously reported \cite{xu2005synthesis, sankar2014crystal, derakhshan2007electronic, lin2022electronic}, a detailed understanding of its structural features and their correlation with physical properties remains unclear. Of particular interest is the role of oxygen coordination in modulating these properties, an area that has not been thoroughly explored. In the dimerized chain direction, long Cu–Cu distances lead to antiferromagnetic (AFM) spin dimers, while short Cu–Cu distances do not exhibit such behavior, underscoring the critical role of oxygen-mediated interactions in shaping the system’s magnetic characteristics. These unresolved issues make NCTO a compelling candidate for further study.

This work presents a comprehensive investigation of Na$_2$Cu$_2$TeO$_6$ (NCTO), a low-dimensional magnetic system in polycrystalline form. The study encompasses synthesis, crystal structure, magnetic properties, heat capacity, and microstructural characteristics of NCTO.

\section{Experimental Details}

A polycrystalline sample of Na$_2$Cu$_2$TeO$_6$ was synthesized using the solid-state reaction method~\cite{sankar2014crystal}. A stoichiometric mixture of Na$_2$CO$_3$ (99.9\%), CuO (99.99\%), and TeO$_2$ (99.99\%) was sintered at 800~$^\circ$C in air for 24 hours with several intermediate grindings at slow heating and cooling rates. The powder X-ray diffraction pattern was recorded using Cu K$\alpha$ radiation at room temperature, employing a D2 Phaser from M/s. Bruker, Germany. Data were collected over a $2\theta$ range of 10$^\circ$--90$^\circ$ with an increment of 0.020$^\circ$. The measured diffraction patterns were analyzed using the Rietveld refinement technique, employing the \textsc{FullProf} suite. 

For temperature-dependent magnetization and isothermal magnetization measurements, a commercial physical property measurement system (helium re-liquefier-based 9~T PPMS and 14~T Dynacool PPMS from M/s. Quantum Design, USA) in Vibrating Sample Magnetometer (VSM) mode was utilized. The dc magnetization measurements were carried out over the temperature range of 2.5--350~K in zero-field-cooled (ZFC) and field-cooled (FC) conditions under applied magnetic fields of 1000~Oe and 5000~Oe. Additionally, isothermal magnetization was studied at 2.5, 10, 50, 100, and 300~K under fields up to 9~T. Specific heat measurements were performed using the two-tau method. The sample was loaded into the specific heat module of the PPMS, and specific heat characterization in the temperature range of 2~K to 310~K allowed detailed thermal property analysis across a wide range of temperatures. The addenda were collected separately and subtracted during the specific heat measurements.

 UV–Vis absorption spectra were recorded using an MS UV Plus double beam spectrophotometer (Motras Scientific Instruments Pvt. Ltd., India), equipped with a 1200 lines/mm blazed holographic grating and dual Hamamatsu silicon photodiodes in a Czerny–Turner configuration, with a variable spectral resolution of 0.1–4.0 nm; measurements were carried out over a wavelength range of 200–1100 nm using pre-aligned halogen and deuterium lamps. Neutron powder diffraction (NPD) measurements were carried out at multiple temperatures (3~K, 10~K, 15~K, 100~K, 150~K, 200~K, 250~K, and 300~K) on the focusing crystal-based powder diffractometer (PD-3) and position-sensitive detector (PSD)-based powder diffractometer at the Dhruva reactor, BARC, Trombay, using neutrons with a wavelength of $\lambda = 1.48~\text{\AA}$ \cite{pramana2008}. The powder samples were packed in a vanadium can with an inner diameter of $\sim$6~mm, and data was recorded over the $2\theta$ range of 6$^\circ$ to 120$^\circ$. The NPD profiles were analyzed for crystallographic structures by fitting the observed patterns to the profiles calculated using the Rietveld refinement method via the \textsc{FullProf} suite program \cite{RODRIGUEZCARVAJAL199355}.

\section{Results and Discussions}

\subsection{Crystal Structure}

\begin{figure}[htbp]
    \centering
	\includegraphics[width=0.5\textwidth, angle= 0]{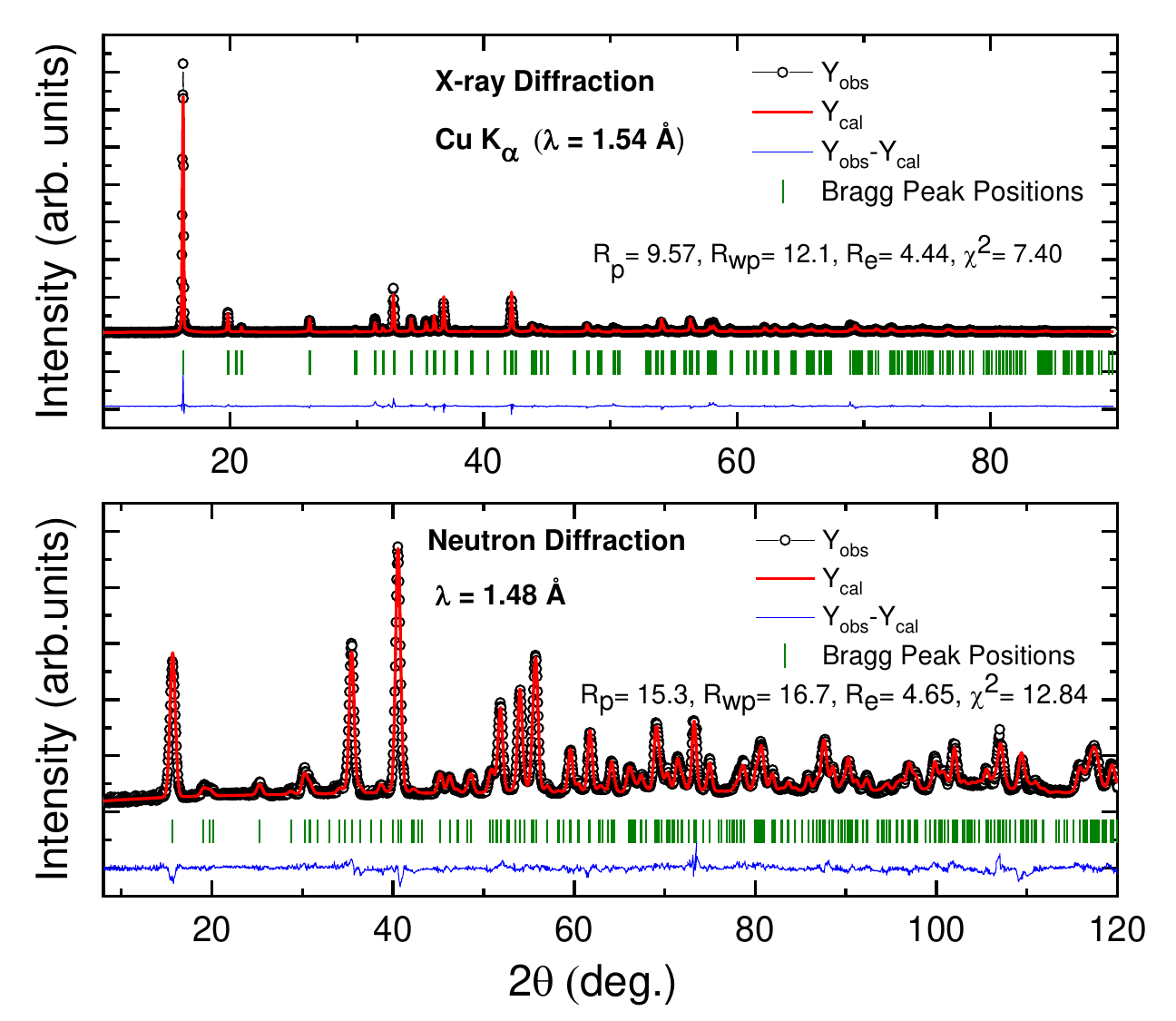}	
	\caption{Experimentally observed (circles) and calculated (solid
line through the data points) (a) x-ray and (b) neutron diffraction
patterns (intensity vs 2 Theta) for Na$_2$Cu$_2$TeO$_6$ at room
temperature. The difference between observed and calculated patterns
is shown by the solid lines at the bottom of each panel. The vertical
bars indicate the positions of allowed nuclear Bragg peaks. Goodness of fitting parameters along with Chi$^2$ values are shown in both the figures.} 
	\label{fig_xray_neutron}%
\end{figure}

The crystal structure of Na$_2$Cu$_2$TeO$_6$ was determined by X-ray and neutron diffraction at room temperature. The Rietveld refinement results, shown in Figs.\ref{fig_xray_neutron} (a) and (b), indicate that the observed diffraction patterns align with the calculated ones. For X-ray diffraction ($\lambda = 1.54 \, \text{\AA}$), the refinement yielded $R_p = 9.57\%$, $R_{wp} = 12.10\%$, $R_e = 4.44\%$, and $\chi^2 = 7.40$, demonstrating a satisfactory fit. For neutron diffraction ($\lambda = 1.48 \, \text{\AA}$), the refinement gave $R_p = 15.30\%$, $R_{wp} = 16.70\%$, $R_e = 4.65\%$, and $\chi^2 = 12.84$. These values are consistent with the greater sensitivity of neutron diffraction to light elements, such as oxygen.

XRD and neutron diffraction refinement was conducted without including any secondary phases, to achieve refinement convergence. This approach indicates the absence of detectable secondary phases in the sample, affirming its phase purity.

\renewcommand{\thetable}{\arabic{table}}
\setcounter{table}{0}
\begin{table*}[htbp!]
\centering
\label{tab:atomic_positions}
\begin{tabular}{lccccc}
\hline
\textbf{Atoms} & \textbf{Wyckoff Position} & \textbf{x/a} & \textbf{y/b} & \textbf{z/c} & \textbf{Occ.} \\ \hline
Cu             & 4g                        & 0            & 0.6647  & 0            & 0.5          \\
Te             & 2a                        & 0            & 0            & 0            & 0.250        \\
O1             & 8j                        & 0.1934  & 0.1638  & 0.2145  & 1            \\
O2             & 4i                        & 0.7562  & 0            & 0.1594  & 0.5          \\
Na             & 4h                        & 0            & 0.1923  & 0.5          & 0.5          \\ \hline
\end{tabular}
\caption{The Rietveld refined lattice constants, fractional atomic
coordinates, and isotropic thermal parameters (Biso) for Na$_2$Cu$_2$TeO$_6$ at room temperature. “Occ.” stands for site occupancy.}
\end{table*}

\begin{table*}[htbp]
\centering
\label{tab:Bond_distances}
\begin{tabular}{lclc}
\hline
\textbf{Atomic Bond} & \textbf{Bond Lengths (Å)} & \textbf{Atomic Bond} & \textbf{Bond Lengths (Å)} \\ \hline
Te-O1               & 1.843(4)                    & Cu-Cu               & 2.393(4)                    \\
Te-O2               & 1.964(3)                    & Cu-Cu               & 3.214(6)                    \\
Cu-O1               & 1.828(6)                    & Na-O1               & 2.464(3)                    \\
Cu-O2               & 1.902(3)                    & Na-O1               & 2.439(7)                    \\
Cu-O1               & 2.500(3)                    & Na-O2               & 2.659(4)                    \\ \hline
\end{tabular}
\caption{Selected Bond Distances (Å) for Na$_2$Cu$_2$TeO$_6$}
\end{table*}

\begin{table}[htbp]
\centering
\label{tab:bond_angles}
\begin{tabular}{lc}
\hline
\textbf{Atomic Bond Angles} & \textbf{Angle (°)} \\ \hline
(\text{Cu})-(\text{Cu})-(\text{Te}) & 179.96(18) \\
(\text{Cu})-(\text{Cu})-(\text{O1}) & 138.6(3)   \\
(\text{Cu})-(\text{Cu})-(\text{O1}) & 89.82(17)  \\
(\text{Cu})-(\text{Cu})-(\text{O2}) & 44.16(15)  \\
(\text{Te})-(\text{Cu})-(\text{O1}) & 41.36(14)  \\
(\text{Te})-(\text{Cu})-(\text{O1}) & 90.18(13)  \\
(\text{Te})-(\text{Cu})-(\text{O2}) & 135.84(12) \\
(\text{O1})-(\text{Cu})-(\text{O1}) & 82.72(19)  \\
(\text{O1})-(\text{Cu})-(\text{O1}) & 90.42(20)  \\
(\text{O1})-(\text{Cu})-(\text{O2}) & 163.3(2)   \\
(\text{O1})-(\text{Cu})-(\text{O2}) & 96.8(2)    \\ \hline
\end{tabular}
\caption{Selected Bond Angles for Na$_2$Cu$_2$TeO$_6$}
\end{table}

\begin{table*}[htbp]
\centering
\label{tab:bondvalencesum}
\begin{tabular}{lcccc}
\hline
\textbf{Atom} & \textbf{Bonding Atom(s)} & \textbf{Bond Distance (Å)} & \textbf{Bond-Valence (S)} & \textbf{Coordination Number} \\ \hline
Cu            & O2                       & 1.9864(32)                & 0.609(5)                  & 4                            \\
Te            & O2                       & 1.9695(41)                & 0.456(5)                  & 6                            \\
O2            & Cu, Te                   & 1.9864(32), 1.9695(41)    & 0.609(5), 0.456(5)        & 4                            \\ \hline
\end{tabular}
\caption{Bond-Valence Analysis for Selected Atoms in Na$_2$Cu$_2$TeO$_6$}
\end{table*}

Rietveld refinement confirms that Na$_2$Cu$_2$TeO$_6$ crystallizes in the monoclinic space group $C2/m$ (s.g. \#12). Neutron diffraction yields refined lattice parameters: $a = 5.7024(3) \, \text{\AA}$, $b = 8.6567(4) \, \text{\AA}$, $c = 5.9389(3) \, \text{\AA}$, and $\beta = 113.739(4)^\circ$. Oxygen ions occupy Wyckoff positions O1 (8j) and O2 (4i), while Cu (4g), Te (2a), and Na (4h) reside at distinct Wyckoff sites (Table \ref{tab:atomic_positions}). Low $R_p$, $R_{wp}$, and $\chi^2$ values confirm the reliability of the model, with neutron refinement indicating phase purity.

X-ray and neutron diffraction study reveals a layered Na$_2$Cu$_2$TeO$_6$ structure composed of Cu$_2$TeO$_6$ units separated by Na ions. Cu atoms form CuO$_4$ plaquettes, while Te forms TeO$_6$ octahedra, alternating to establish the layered framework as seen in Fig.\ref{fig_CrystalStructure}a. Similarly, Fig. \ref{fig_CrystalStructure}b highlights Cu$_2$O$_6$ dimers with distorted non-planar plaquettes, showing distinct Cu-O1 and Cu-O2 bond lengths. Unlike Na$_2$Co$_2$TeO$_6$, Na$_2$Cu$_2$TeO$_6$ lacks honeycomb symmetry due to preferential Cu-O bonding in square-planar CuO$_2$ units, resulting in unequal Cu-O bond lengths: Cu-O1 = 1.977(4) \AA, Cu-O2 = 1.986(4) \AA (Fig. 1b). Cu-Cu distances vary from 2.393 \AA~to 3.214 \AA~(Table \ref{tab:Bond_distances}), where shorter distances correspond to stronger interactions. Cu-O bond lengths span 1.828 \AA~to 2.500 \AA, signifying strong Cu-O interactions, while Na-O bond lengths (2.439 \AA~to 2.659 \AA) indicate weaker interactions due to Na's larger ionic radius and lower charge density.

Bond angles (Table \ref{tab:bond_angles}) further elucidate the structure. The Cu-Cu-Te bond angle is nearly linear at 179.96$^\circ$, indicating minimal Cu-Cu distortion. Cu-O bond angles vary significantly: Cu-O1-Cu = 138.6$^\circ$ (nearly ideal), whereas Cu-O2-Cu = 89.82$^\circ$ (suggesting angular strain). Te-O-Cu angles range from 41.36$^\circ$ to 135.84$^\circ$, reflecting Te's complex coordination. Bond-valence calculations confirm stronger Cu-O compared to Te-O bonding, consistent with Cu’s higher electropositivity (Table \ref{tab:bondvalencesum}). 

These results emphasize the interplay between bond lengths, angles, and coordination geometries in shaping the structural and functional properties of Na$_2$Cu$_2$TeO$_6$.


The Cu$_2$TeO$_6$ layers in Na$_2$Cu$_2$TeO$_6$ consist of edge-sharing TeO$_6$ and CuO$_6$ octahedra. Each CuO$_6$ octahedron is axially elongated due to the Jahn-Teller distortion of Cu$^{2+}$ ions, forming Cu$_2$O$_{10}$ dimers via edge-sharing. The Cu$^{2+}$ ions arrange in a distorted honeycomb lattice within the Cu$_2$TeO$_6$ layers, with Te$^{6+}$ ions occupying the centers of the Cu$^{2+}$ hexagons. The spin exchange interactions in Na$_2$Cu$_2$TeO$_6$ are primarily governed by three exchange pathways: \( J_1 \), \( J_2 \), and \( J_3 \), corresponding to distinct Cu$^{2+}$-Cu$^{2+}$ linkages within the Cu$_2$TeO$_6$ layer.

The nearest-neighbor interactions \( J_1 \) and \( J_3 \) are superexchange (SE) interactions mediated by Cu-O-Cu linkages, while the next-nearest-neighbor interaction \( J_2 \) is a super-superexchange (SSE) interaction mediated by Cu-O...O-Cu linkages. The bond lengths and angles associated with these exchange paths directly influence the magnetic interactions in Na$_2$Cu$_2$TeO$_6$.

The geometrical parameters of the spin exchange paths \( J_1 \), \( J_2 \), and \( J_3 \) are summarized in Table \ref{tab:spin_exchange_paths_na2}, detailing the bond lengths and angles that govern the magnetic coupling in the system.

\begin{table}[htbp]
\centering
\label{tab:spin_exchange_paths_na2}
\begin{tabular}{l c}
\hline
\textbf{Spin Exchange Path} & \textbf{Na$_2$Cu$_2$TeO$_6$} \\
\hline
J1: Cu-Cu Distance & 5.802(7) \, Å \\
J1: Cu-O1-Cu Angle & 128.81(12)° \\
\hline
J2: Cu-Cu Distance & 2.850(7) \, Å \\
J2: Cu-O2-Cu Angle & 91.68(19)° \\
\hline
J3: Cu-Cu Distance & 3.208(4) \, Å \\
J3: O1-O1 Distance & 3.231(6) \, Å \\
J3: Cu-O1/O2-Cu Angle & 89.58(13)°, 62.72(7)° \\
\hline
\end{tabular}
\caption{Geometrical Parameters Associated with the Spin Exchange Paths J1, J2, and J3 of Na$_2$Cu$_2$TeO$_6$}
\end{table}

In the crystal structure of Na$_2$Cu$_2$TeO$_6$, the spin exchange interactions are illustrated in Fig.~\ref{fig_CrystalStructure}(c), where the \( J_1 \), \( J_2 \) (along the chain), and \( J_3 \) (between the chains) paths are highlighted. The corresponding geometrical parameters are summarized in Table~\ref{tab:spin_exchange_paths_na2}. Specifically, the \( J_1 \) interaction path corresponds to a Cu-Cu distance of 5.802(7) Å with a Cu-O$_1$-Cu angle of 128.81(12)°. The \( J_2 \) path exhibits a shorter Cu-Cu distance of 2.850(7) Å and a Cu-O$_2$-Cu angle of 91.68(19)°. In contrast, the interchain \( J_3 \) interaction has a Cu-Cu distance of 3.208(4) Å and an O$_1$-O$_1$ distance of 3.231(6) Å, with Cu-O$_1$/O$_2$-Cu angles of 89.58(13)° and 62.72(7)°, respectively.

Despite the longer Cu-Cu separation in the \( J_1 \) path, it exhibits stronger coupling than \( J_2 \) due to the Cu-O···O-Cu super-superexchange mechanism, which induces antiferromagnetic (AF) interactions between Cu$^{2+}$ spins. Conversely, the \( J_2 \) couplings are ferromagnetic (FM), attributed to the nearly 90° Cu-O$_2$-Cu angle, which facilitates orbital overlap, as indicated in Table~\ref{tab:spin_exchange_paths_na2}. These results are consistent with previous findings reported in \cite{gao2020weakly}.

\subsection{UV-Visible Spectroscopy Study}

The optical absorption behavior of Na$_2$Cu$_2$TeO$_6$ was examined using UV-Visible spectroscopy. Figure~\ref{fig:uvvis_combined}a shows the absorbance spectrum as a function of wavelength. A sharp increase in absorbance in the UV region\begin{figure}[H]
    \centering
    \includegraphics[width=0.5\textwidth]{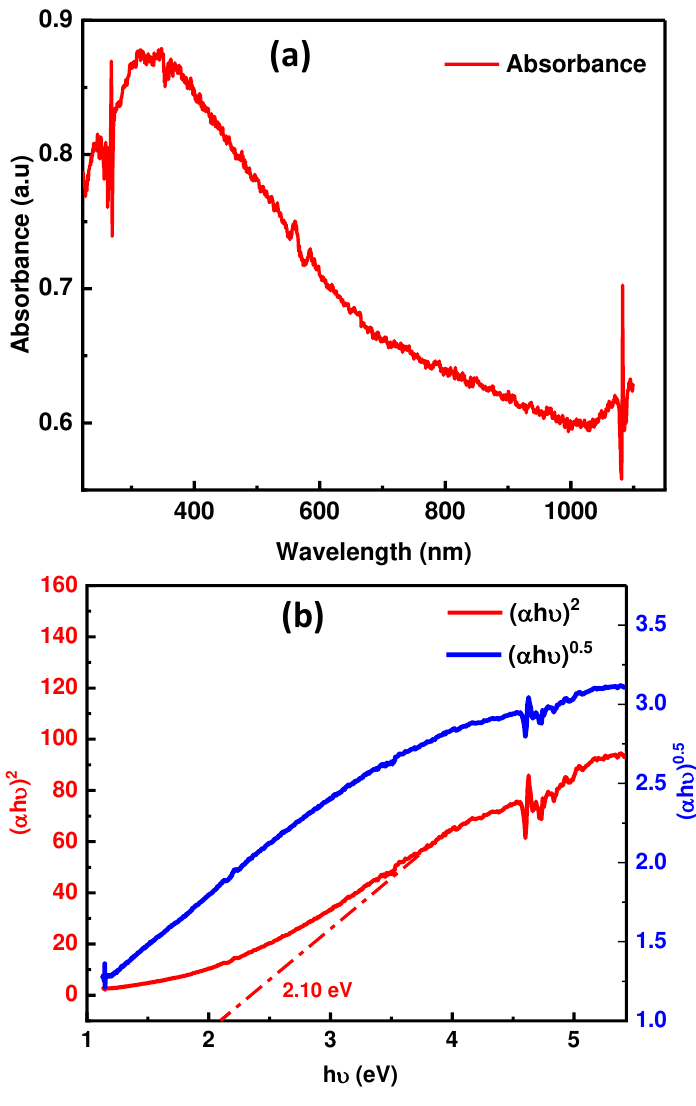} 
    \caption{(a) UV-Visible absorbance spectrum of \ce{Na2Cu2TeO6} powder suspended in liquid. (b) Tauc plots for direct allowed transition (\((\alpha h\nu)^2\), red) and indirect allowed transition (\((\alpha h\nu)^{0.5}\), blue). A direct band gap of 2.10 eV is estimated from the red curve.}
    \label{fig:uvvis_combined}
\end{figure}indicates the onset of electronic transitions across a wide band gap.

To estimate the optical band gap, Tauc plots were constructed assuming both direct allowed (\(n = 2\)) and indirect allowed (\(n = 0.5\)) transitions using the Tauc relation:

\begin{equation}
(\alpha h\nu)^n = A(h\nu - E_g)
\end{equation}

where $h\nu$ is the photon energy, $E_g$ is the optical band gap, $\alpha$ is the absorption coefficient (in arbitrary units due to unknown thickness), and $n$ depends on the type of transition.

As shown in Figure~\ref{fig:uvvis_combined}b, the Tauc plot for the direct transition (\((\alpha h\nu)^2\)) exhibits a clear linear region, and extrapolation of this segment yields a direct band gap of approximately \textbf{2.10 eV}. In contrast, the Tauc plot for the indirect transition (\((\alpha h\nu)^{1/2}\)) does not show a well-defined linear portion or a clear x-intercept, making it unreliable for band gap determination. This suggests that Na$_2$Cu$_2$TeO$_6$ is more likely to exhibit a direct allowed transition under the present experimental conditions.

\subsection {Magnetic Susceptibility}

\begin{figure}[H]
    \centering
    \includegraphics[width=0.5\textwidth]{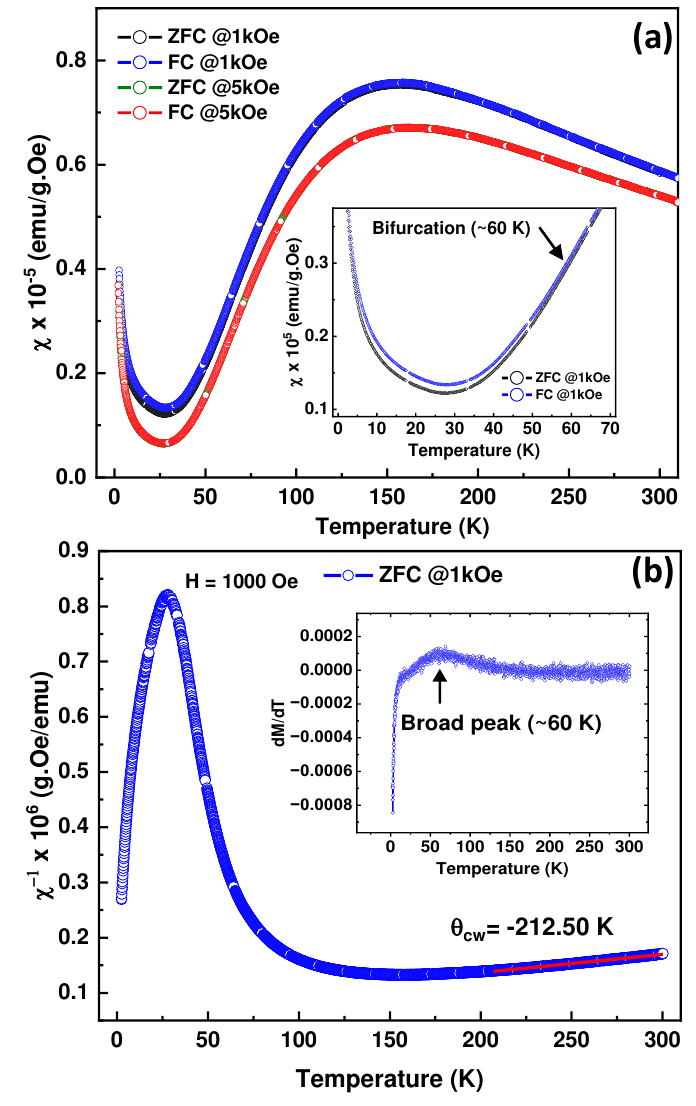}
    \caption{(a) Temperature-dependent magnetization curves measured in Zero-Field Cooled (ZFC) and Field-Cooled (FC) modes under applied magnetic fields of 1000 Oe and 5000 Oe. The inset highlights the bifurcation observed in the ZFC-FC curves at 1000 Oe. (b) Temperature-dependent inverse dc susceptibility (M/H) curves under ZFC and FC conditions with a 1000 Oe applied magnetic field. The solid red lines represent the Curie-Weiss fit to the data. The inset shows the transition occurring around 60 K.} 
    \label{fig_MT_XT}
\end{figure}

Temperature-dependent magnetization measurements of Na$_2$Cu$_2$TeO$_6$ were conducted in Zero-Field-Cooled (ZFC) and Field-Cooled (FC) modes at 1000 Oe and 5000 Oe, as shown in Fig. \ref{fig_MT_XT}(a). The data exhibit characteristics of frustrated magnetism, where competing interactions prevent long-range magnetic ordering.

The ZFC and FC magnetization curves show a bifurcation onset at approximately 60~K (see inset of Fig. \ref{fig_MT_XT}(a)), which coincides with the broad peak observed in the $\frac{dM}{dT}$ curve near the same temperature (see inset of Fig. \ref{fig_MT_XT}(b)). This concurrence strongly indicates the onset of short-range magnetic correlations or frustration-driven phenomena. In particular, the broad nature of the peak reflects the gradual development of a magnetically disordered state, possibly related to spin freezing phenomena commonly observed in frustrated low-dimensional systems \cite{Gardner2009, sankar2014crystal}.

In the ZFC mode, the magnetization increases with decreasing temperature, showing a broad maximum around 160~K, typical of short-range interactions. In the FC mode, the magnetization exhibits an upturn near 30~K that persists down to 2.5~K, a hallmark of frustrated systems where competing interactions inhibit the establishment of long-range order \cite{sankar2014crystal, MURTAZA20211601}.

At 5000 Oe, the bifurcation between ZFC and FC curves diminishes, and the broad maximum near 160~K becomes more extended, indicating suppression of frustration effects by the applied field. This stabilization favors the alignment of magnetic moments \cite{Ramirez1994-tn, Greedan2001, Schiffer1996, Gingras2014, Feng2010}. The upturn at 30~K seen in the 1000 Oe data suggests a magnetic transition possibly related to a magnetically disordered or frozen state, which is less pronounced at 5000 Oe, further supporting field-induced suppression of competing interactions.

The inverse DC magnetic susceptibility, $\chi^{-1}(T)$, shown in Fig. \ref{fig_MT_XT}(b), reveals a linear paramagnetic region between 210~K and 300~K, confirming the paramagnetic nature of the material at high temperatures. The data were fitted to the Curie-Weiss equation:

\begin{equation}
\chi^{-1}(T) = \frac{T}{C} + \theta_{CW}
\end{equation}

where $C$ is the Curie constant and $\theta_{CW}$ is the Curie-Weiss temperature, providing insight into the nature of the magnetic interactions. The linear fit yields a Curie-Weiss temperature of $\theta_{CW} = -212.5$~K, indicative of dominant antiferromagnetic interactions in Na$_2$Cu$_2$TeO$_6$. The negative $\theta_{CW}$ suggests a frustrated magnetic ground state, where competing interactions result in short-range correlations instead of long-range ordering. From the fit, the Curie constant is $C = 0.0032(1)$ emu K/mol·Cu·Oe, and the effective magnetic moment, $\mu_{\text{eff}}$, is given by:

\begin{equation}
\mu_{\text{eff}} = \sqrt{8C} \, \mu_B
\end{equation}

The calculated effective moment is \( \mu_{\text{eff}} = 0.16 \, \mu_B \), significantly smaller than the expected value of \( \mu_{\text{eff}} = 1.73 \, \mu_B \) for a Cu$^{2+}$ ion with spin $S = 1/2$ and $g \approx 2$. This reduction is characteristic of frustrated systems, where competing interactions prevent full alignment of magnetic moments, thereby suppressing the effective magnetic moment.

The $\frac{dM}{dT}$ curve, shown in the inset of Fig.~6(b), exhibits a broad peak near 60~K, indicative of a magnetic phase transition. This feature, coinciding with the ZFC-FC bifurcation, marks the onset of short-range magnetic correlations likely arising from low-dimensional interactions and magnetic frustration. Below this temperature, the system transforms toward a magnetically disordered state, possibly related to spin freezing phenomena commonly seen in frustrated magnets \cite{Gardner2009, sankar2014crystal}.

The Pearson's $R$ value of 0.99659 confirms a strong linear correlation between $\chi^{-1}$ and $T$, supporting the Curie-Weiss behavior. The $R^2$ value of 0.99319 and adjusted $R^2$ of 0.99318 indicate that the linear model accounts for approximately 99.3\% of the data variance, with minimal overfitting. The residual sum of squares, $7.76324 \times 10^8$, reflects small deviations from the fit, which are insignificant considering the data range.

\subsection {Isothermal Magnetization}
We conducted isothermal magnetization (M-H) studies to investigate the magnetic behavior of Na$_2$Cu$_2$TeO$_6$. At high temperatures (300 K and 100 K), the M-H curves show linear field dependence across the entire applied field range, typical of the paramagnetic regime, where magnetic moments respond independently to the field. 

This linearity suggests negligible magnetic interactions at elevated temperatures, as the magnetization is directly proportional to the applied field \cite{Kittel2005}. Linear fitting of the M-H data yields high-quality fits with $R^2 = 1$ at both 300 K and 100 K, confirming the paramagnetic nature of the material. \begin{figure}[H]
    \centering
    \includegraphics[width=0.5\textwidth]{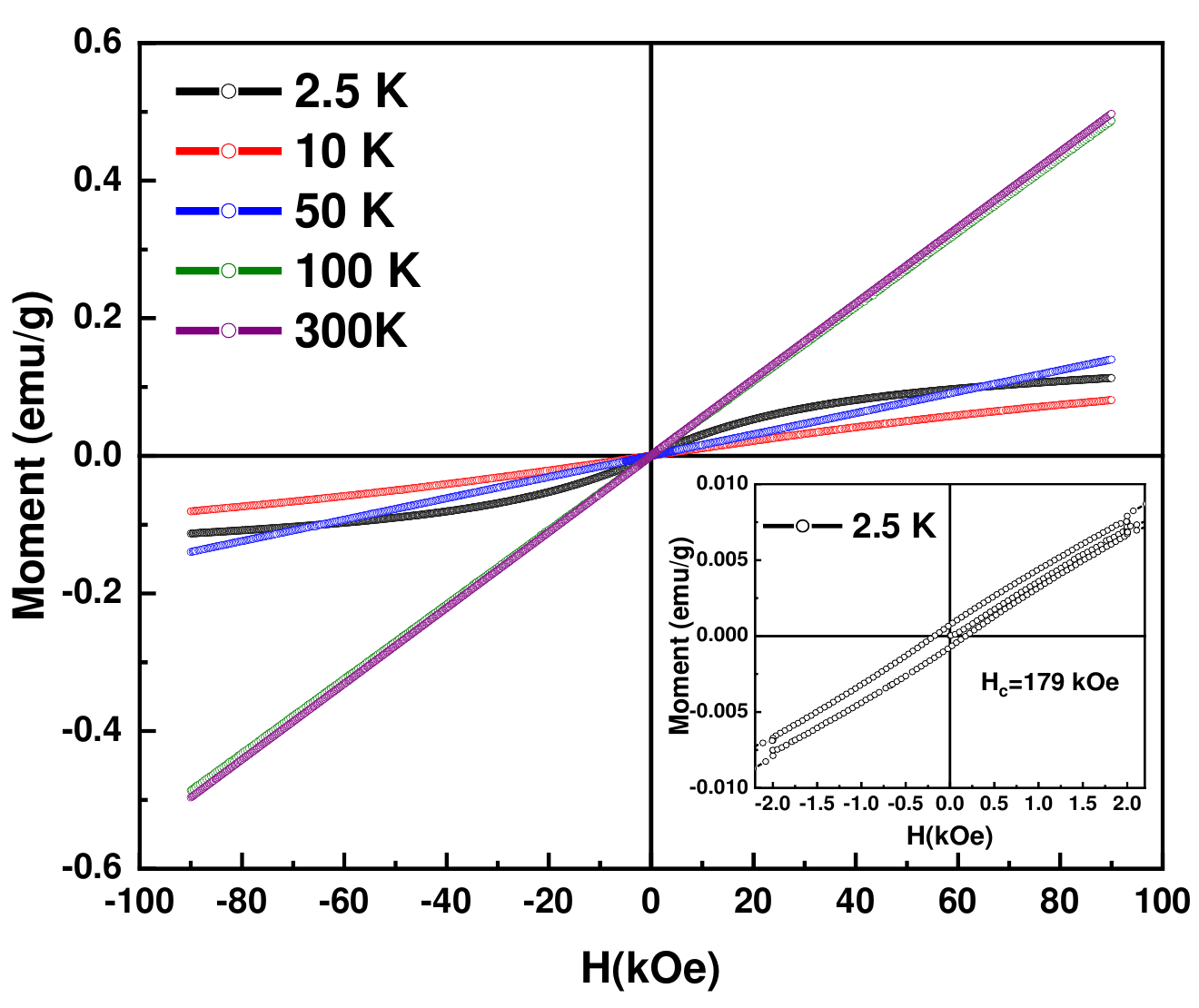}
    \caption{The isothermal magnetization of Na2Cu2TeO6 as a
function of magnetic field measured at 2.5 K, 10 K, 50 K, 100 K and 300 K.  The inset shows the hysteresis loop with coercive field of 179 kOe.} 
    \label{fig_MvsH}
\end{figure} The linear fitting parameters, including slope and intercept, along with the key observations for the M-vs-H data at lower temperatures (100 K, 50 K, 10 K, and 2.5 K), are summarized in Table 7. At 300 K, the intercept and slope values of $6.499 \times 10^{-4}$ emu/g and $5.514 \times 10^{-6}$ emu/g$\cdot$Oe, respectively, correspond to pure paramagnetic behavior, while at 100 K, the intercept and slope values slightly decrease to $-0.1853 \times 10^{-4}$ emu/g and $5.406 \times 10^{-6}$ emu/g$\cdot$Oe.

At 50 K, a slight deviation from linearity at higher fields signals the onset of weak low-temperature magnetic correlations, indicating the increasing influence of exchange interactions. This results in a gradual transition from paramagnetism to weakly correlated behavior, likely driven by short-range interactions. The slope reduces to $1.552 \times 10^{-6}$ emu/g$\cdot$Oe, while the intercept shows a small positive value of $0.0998 \times 10^{-4}$ emu/g, suggesting the emergence of short-range correlations as temperature decreases. A similar trend is observed in CuTeO$_4$, where short-range correlations dominate at low temperatures \cite{Hasan2024Quasi}. 

\begin{table*}[htbp!]
\centering
\resizebox{\textwidth}{!}{
\begin{tabular}{lcccccl}
\hline
\textbf{Temperature (K)} & \textbf{Intercept (10$^{-4}$ emu/g)} & \textbf{Slope (10$^{-6}$ emu/g·Oe)} & \textbf{R$^2$} & \textbf{Key Observations} \\ \hline
300 & 6.499  & 5.514   & 1.000 & Purely paramagnetic behavior. \\
100 & -0.1853 & 5.406 & 1.000 & Predominantly paramagnetic response. \\
50  & 0.0998 & 1.552  & 0.99999 & Onset of weak low-temperature correlations. \\
10  & 0.0998 & 1.552  & 1.000 & Signs of field-induced effects. \\
2.50 (upto 1 T) & 27.19 & 1.560 & 0.961 & Strong low-field correlations. \\
2.5 (6–9 T) & 660.8 & 0.5252 & 0.997 & Partial magnetic saturation observed. \\ \hline
\end{tabular}
}
\caption{Linear Fitting Parameters for M-vs-H Data at Different Temperatures}
\label{tab:linear_fitting_params}
\end{table*}

At 2.5 K, the M-H curve deviates significantly from linearity, especially in the 6–9 T range, where partial magnetization saturation is observed. The fitting parameters (Table \ref{tab:linear_fitting_params}) reveal strong low-field correlations up to 1 T, with partial saturation occurring at higher fields. This deviation suggests a transition to a magnetically correlated state, driven by frustrated interactions and marked by significant anisotropy. These observations align with those in other frustrated and spin-dimer systems \cite{Zhou2020Particle, samulon2004}. 

The inset shows a hysteresis loop at 2.5 K with a coercive field of 179 kOe, signaling a transition to a magnetically correlated state. The presence of hysteresis suggests significant magnetic anisotropy and disorder, indicative of frustrated interactions \cite{Lacroix2011}.

\subsection {Specific heat analysis}
Following magnetization studies of Na$_2$Cu$_2$TeO$_6$, which revealed an absence of long-range magnetic ordering \cite{Kittel2005}, specific heat ($C_p$) measurements were performed to probe thermodynamic behavior and explore short-range magnetic correlations in the system. Specific heat was measured up to 310~K, showing a continuous paramagnetic-like increase, devoid of sharp anomalies or plateaus. This behavior corroborates magnetization data, suggesting dominance of short-range magnetic correlations over long-range interactions \cite{Ramirez}.

Low-temperature specific heat data were analyzed using the expression:
\begin{equation}
C_p(T) = \gamma T + \beta T^3,
\end{equation}
where linear term ($\gamma T$) represents electronic contribution and cubic term ($\beta T^3$) arises from lattice vibrations, as described by Debye theory \cite{DebyeModel}. Fitting $C_p/T$ versus $T^2$ in low-temperature regime yielded parameters $\gamma = 3.25$~mJ~mol$^{-1}$~K$^{-2}$ and $\beta = 0.61$~mJ~mol$^{-1}$~K$^{-4}$.

Debye temperature ($\Theta_D$) was calculated using the relation:
\begin{equation}
\Theta_D = \left(\frac{12\pi^4 R}{5\beta}\right)^{1/3},
\end{equation}
where $R = 8.314$~J~mol$^{-1}$~K$^{-1}$ is gas constant. The calculated $\Theta_D = 147.67$~K provides insight into vibrational dynamics of the lattice, reflecting strong interatomic bonding and rigid lattice vibrations \cite{AshcroftMermin}. 

Absence of phase transition anomalies in specific heat data, even at lowest measured temperatures, further emphasizes role of short-range magnetic interactions in this material. These results complement magnetic characterization, reinforcing hypothesis that Na$_2$Cu$_2$TeO$_6$ is highly frustrated magnetic system, potentially hosting exotic magnetic ground states of significant scientific interest \cite{Anderson1987}.

\begin{figure}
	\centering 
	\includegraphics[width=0.5\textwidth, angle= 0]{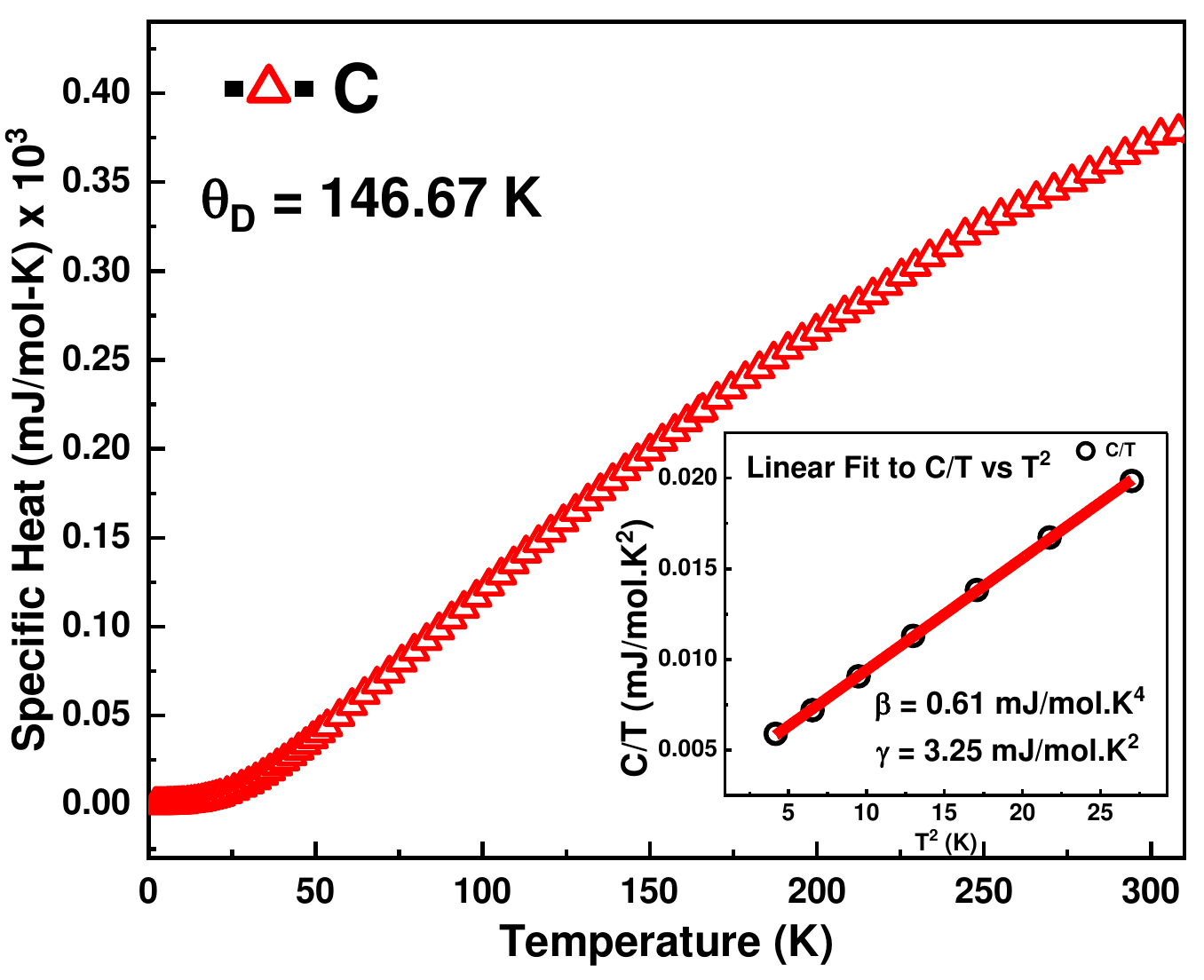}	
	\caption{Specific heat (C) as a function of temperature (T) for Na$_2$Cu$_2$TeO$_6$. The inset shows a C/T vs. T$^2$ plot with a linear fit, used to calculate the Debye temperature ( \$theta$_D$= 146.67 K).} 
	\label{fig_Cv}%
\end{figure}

\subsection{Temperature Dependent Neutron Diffraction}

The temperature-dependent neutron diffraction data for Na$_2$Cu$_2$TeO$_6$ was analyzed over a temperature range of 3 K to 300 K to investigate the evolution of the crystal structure. The unit cell parameters, including the lattice constants \(a\), \(b\), and \(c\), and the angle \(\beta\), show minimal variations with temperature. The unit cell volume \(V\) exhibits a gradual increase from 266.93 Å$^3$ at 3 K to 268.37 Å$^3$ at 300 K, consistent with the typical thermal expansion behavior of ionic materials. 

\begin{figure*}[ht]
	\centering 
	\includegraphics[width=1.0\textwidth, angle=0]{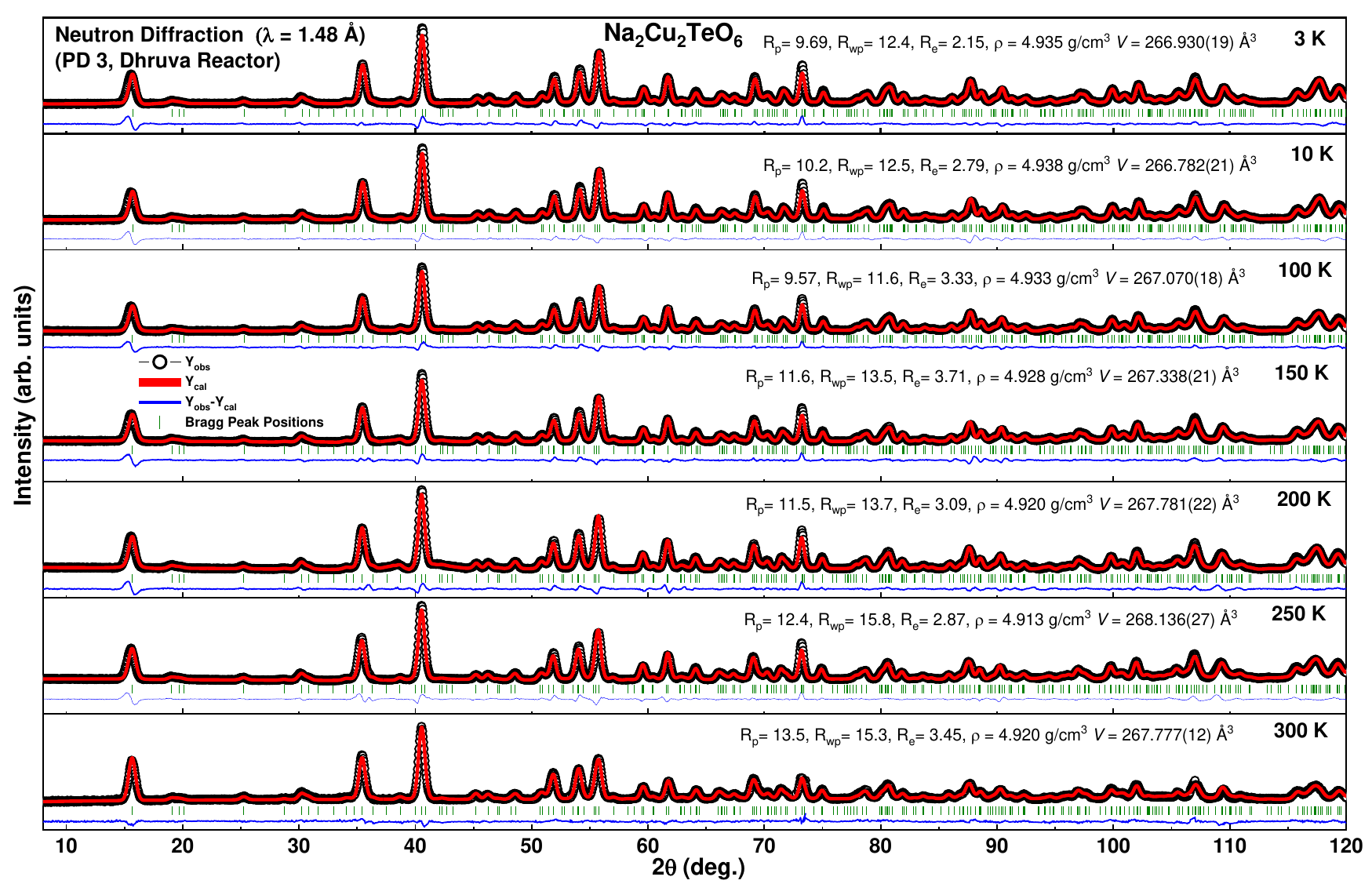}	
	\caption{Temperature-dependent Rietveld-analyzed neutron diffraction patterns from 3 K to 300 K} 
	\label{fig_Neutron_Temp}%
\end{figure*}

This increase can be described by the equation for thermal expansion:
\begin{equation}
V(T) = V_0 (1 + \alpha T)    
\end{equation}

where \(V_0\) represents the unit cell volume at a reference temperature, and \(\alpha\) is the thermal expansion coefficient. These results suggest weak thermal vibrations of atoms, with the volume expansion being relatively small across the measured temperature range.

We also observed that the Cu-O bond lengths (Cu-O1 and Cu-O2) exhibit small temperature-dependent changes. At 3 K, the Cu-O1 bond measures 1.976 Å and the Cu-O2 bond measures 1.992 Å. As the temperature increases to 300 K, the Cu-O1 bond length slightly shortens to 1.975 Å, and the Cu-O2 bond decreases to 1.986 Å. The changes in these bond lengths are minimal, indicating weak spin-lattice coupling, and are consistent with the weak thermal expansion observed in the unit cell volume \cite{Balents2010}.

The absence of magnetic peaks in the diffraction data from 3 K to 300 K suggests that Na$_2$Cu$_2$TeO$_6$ may exhibit a quantum disordered ground state, akin to quantum spin liquids, with no evidence of long-range magnetic ordering.

The scattering equation for neutron diffraction can be described as:

\begin{equation} 
I(\theta) = \sum_{hkl} \left| F_{hkl} \right|^2 \exp\left( -2W \sin^2\theta/\lambda^2 \right)
\end{equation}

where \(I(\theta)\) is the intensity of the scattered neutrons at an angle \(\theta\), \(F_{hkl}\) is the structure factor, and \(W\) is the Debye-Waller factor, which accounts for atomic vibrations. The absence of magnetic peaks in the data further supports the lack of long-range magnetic ordering.

\section{Conclusion}
In this study, we examined the structural and magnetic properties of Na\(_2\)Cu\(_2\)TeO\(_6\). X-ray and neutron diffraction confirmed a monoclinic \(C2/m\) structure with Cu\(^{2+}\) ions forming distorted honeycomb layers. The inequivalent Cu–Cu distances suggest a complex exchange interaction network.

Magnetic susceptibility measurements yielded a large negative Curie-Weiss temperature (\(\theta_{\text{CW}} = -212.5\,\text{K}\)) and a reduced effective moment (\(\mu_{\text{eff}} = 0.16\,\mu_B\)), indicating strong antiferromagnetic interactions and significant quantum fluctuations. A broad maximum near 160 K suggests short-range correlations, while low-temperature specific heat follows a power-law behavior.

Neutron diffraction data down to 3 K revealed no magnetic Bragg peaks, indicating the absence of long-range magnetic order. These results highlight Na\(_2\)Cu\(_2\)TeO\(_6\) as a candidate for further exploration of low-dimensional frustrated magnetism.

\section{Acknowledgement}
S. Patil thanks the UGC-DAE CSR Mumbai Centre, for providing GATE institute fellowship
(PH24S22073392)

\end{document}